\title{\bf Neutrino Mass and Mixing in the Seesaw Playground\footnote{Invited contribution to the Special Issue on Neutrino Oscillation in order
to celebrate the Nobel Prize in Physics 2015 - for publication in Nuclear Physics B}}

\author{ 
  {Stephen F.~King\footnote{\tt king@soton.ac.uk}},
\\
  {\small \it Physics and Astronomy, University of Southampton, Southampton, SO17 1BJ, U.K.}\\
       }

\date{}
\documentclass[11pt]{article}
\pdfoutput=1
\usepackage[margin=2cm]{geometry}
\usepackage{color}
\usepackage{graphicx}
\usepackage{amssymb}
\usepackage{amsmath} 
\usepackage{pifont} 
\usepackage{hyperref}
\usepackage[noadjust]{cite}
\usepackage[center,footnotesize,hang]{subfigure}
\usepackage{multirow}

\usepackage{braket}

\usepackage{mathtools}
\usepackage{amssymb}
\usepackage{amsfonts}
\usepackage[footnotesize]{caption}
\usepackage{color}
\usepackage{braket}
\usepackage{cite}
\usepackage{hyperref}
\usepackage{url}
\usepackage{multirow}
\usepackage{relsize}
\usepackage{fullpage}
\usepackage{makecell}
\usepackage{fullpage}

\newcommand{\pmatr}[1]{\begin{pmatrix} #1 \end{pmatrix}}

\newcommand{\CP}{$\mathcal{CP}$\,}

\begin{document}

\maketitle

\begin{abstract} 
We discuss neutrino mass and mixing in the framework of the classic seesaw mechanism,
involving right-handed neutrinos 
with large Majorana masses, which provides an appealing way to understand
the smallness of neutrino masses.
However, with many input parameters, the seesaw mechanism is in general not predictive.
We focus on natural implementations of the seesaw mechanism, in which large cancellations do not occur,
where one of the right-handed neutrinos is dominantly responsible
for the atmospheric neutrino mass, while a second right-handed neutrino accounts for the solar neutrino mass, leading to an effective two right-handed neutrino model. We discuss recent attempts to predict 
lepton mixing and CP violation within such natural frameworks,
focussing on the Littlest Seesaw and its distinctive predictions.
 \end{abstract}



\section{Introduction}

Although the discovery of neutrino oscillations implying mass and mixing can be regarded as one of the greatest discoveries  
in physics in the last two decades, not least because it provides the only laboratory evidence for physics
beyond the Standard Model (SM), it remains a sobering fact that we still do not know 
the origin of neutrino mass and mixing (for reviews see e.g.\cite{King:2015aea}).
However, at least there seems to be a leading candidate for neutrino mass and mixing, namely the seesaw mechanism involving additional
right-handed neutrinos with heavy Majorana masses \cite{seesaw}.
Although the seesaw mechanism represents an astonishingly elegant 
explanation of the smallness of neutrino mass, it involves many parameters
making quantitative predictions of neutrino mass and mixing challenging,
but not impossible, as we shall discuss.

In this paper we 
focus on natural implementations of the seesaw mechanism in which large cancellations do not occur, where typically one of the right-handed neutrinos is dominantly responsible
for the atmospheric neutrino mass \cite{King:1998jw}, while a second right-handed neutrino accounts for the solar neutrino mass \cite{King:1999mb}. After reviewing the unanswered questions of neutrinos
and lepton mixing, we enter the seesaw playground and discuss recent attempts to try to understand
lepton mixing and CP violation within such natural frameworks, focussing on the Littlest Seesaw with its distinctive predictions.

\begin{figure}[htb]
\centering
\includegraphics[width=0.5\textwidth]{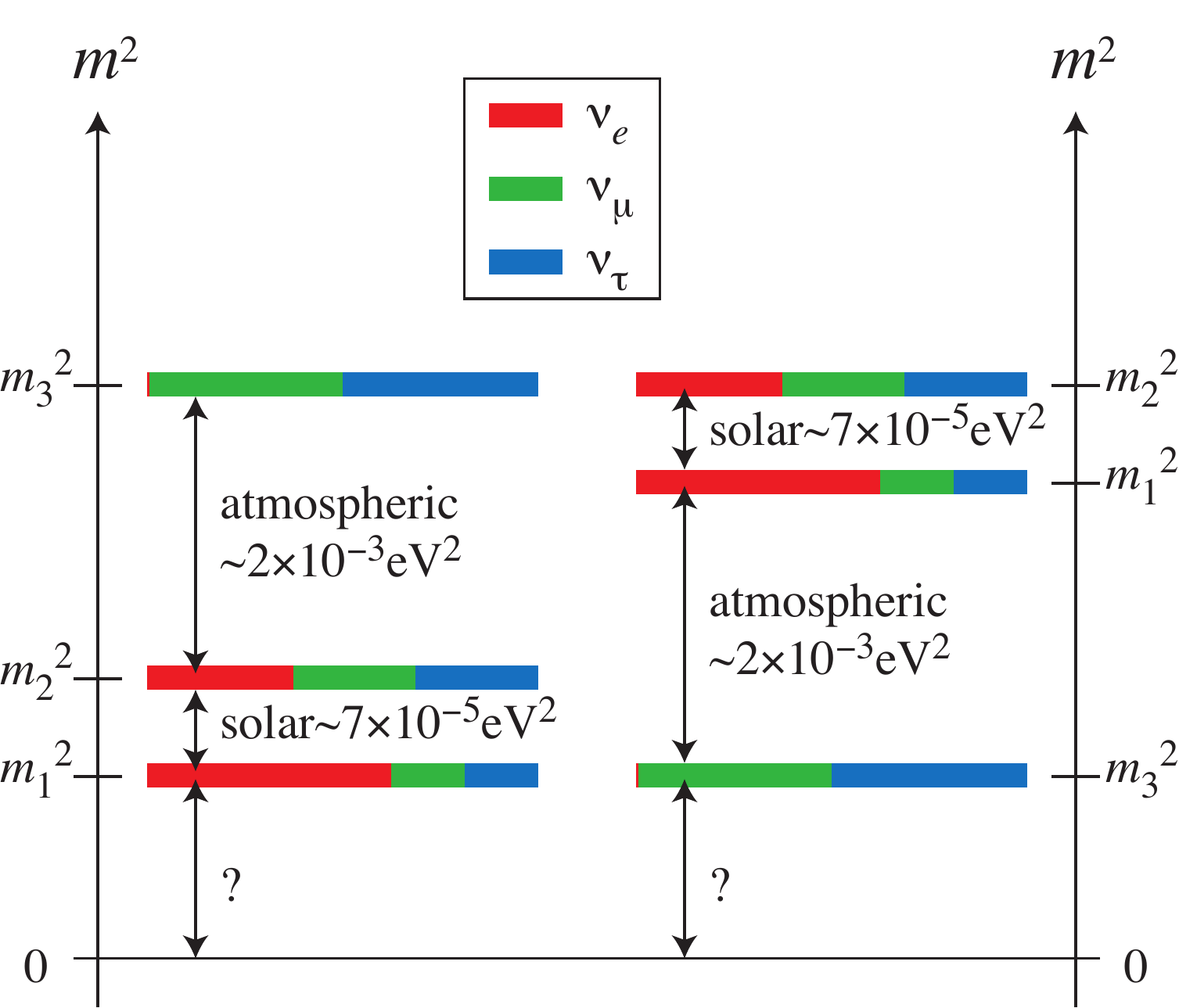}
\caption{\label{mass}\small{The probability that a particular neutrino
mass state $\nu_i$ with mass $m_i$ contains a particular charged lepton mass basis state 
$(\nu_e, \nu_{\mu}, \nu_{\tau})$ is represented by colours.
The left and right panels of the figure are 
referred to as normal or inverted mass squared ordering, respectively,
referred to as NO or IO.
The value of the lightest neutrino mass is presently unknown.
The current best fit values of the mass squared differences are given in 
    \cite{Gonzalez-Garcia:2014bfa,Capozzi:2013csa,Forero:2014bxa}.
    For example, the best fit mass squareds for a normal neutrino mass ordering are
    \cite{Gonzalez-Garcia:2014bfa} are:
   $m_3^2-m_1^2=(2.547\pm 0.047)\times10^{-3}\ {\rm eV}^2$and 
   $m_2^2-m_1^2=(7.50\pm 0.18)\times10^{-5}\ {\rm eV}^2$.}}
\end{figure}

 \section{Unanswered questions of neutrinos}
\begin{figure}[t]
\centering
\includegraphics[width=0.80\textwidth]{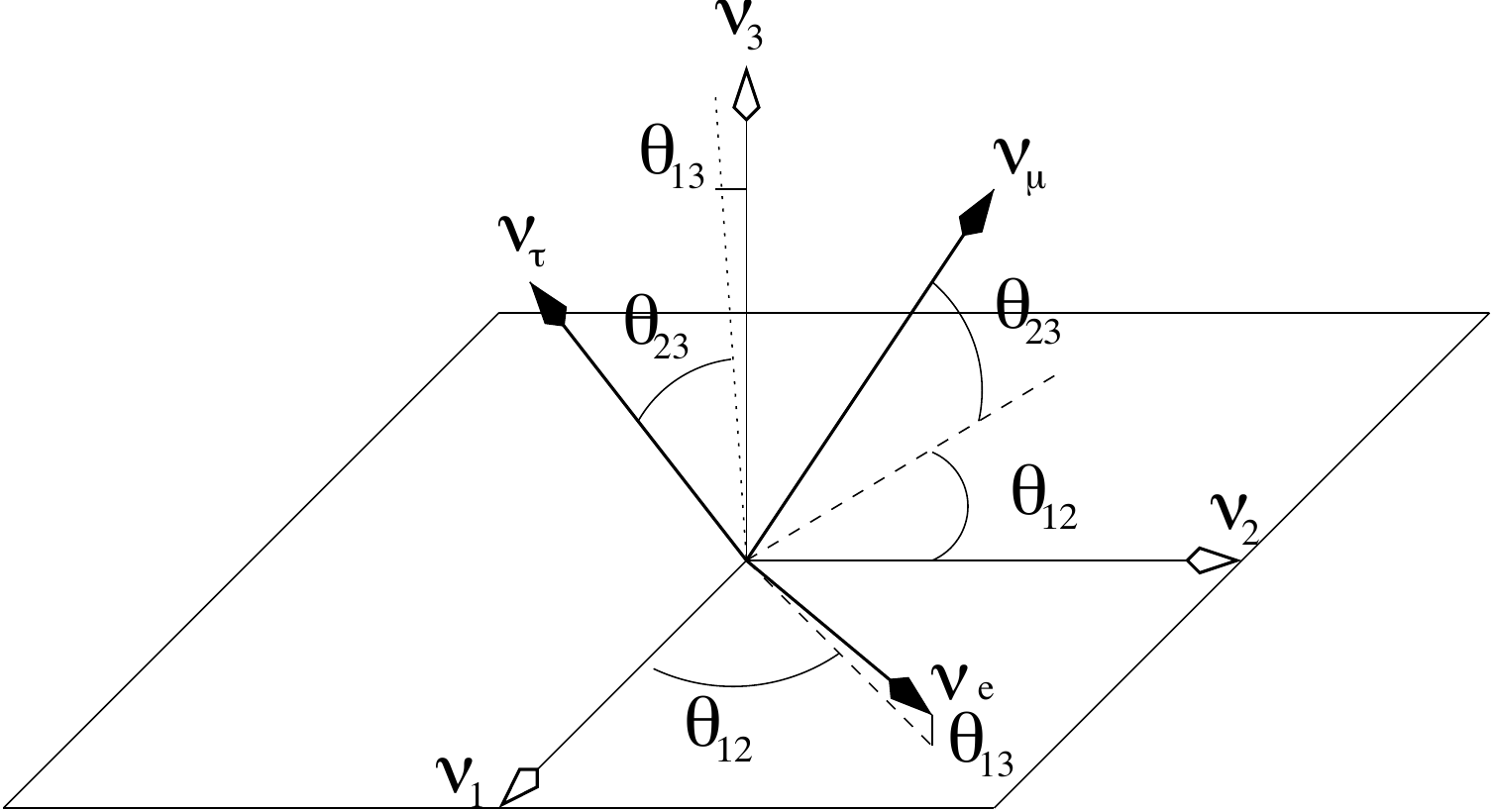}
    \caption{Lepton mixing angles (assuming zero \CP violation) may be represented as Euler angles relating the charged lepton mass basis states $(\nu_e, \nu_{\mu}, \nu_{\tau})$ to the mass eigenstate basis states
    $(\nu_1, \nu_2, \nu_3)$.
    The current bext fit values of the angles are given in 
    \cite{Gonzalez-Garcia:2014bfa,Capozzi:2013csa,Forero:2014bxa}.
    For example, the best fit angles for a normal neutrino mass ordering 
    quoted in \cite{Gonzalez-Garcia:2014bfa} are (in degrees):
    $\theta_{12}=33.48^{+0.78}_{-0.75}$, $\theta_{23}=42.3^{+3.0}_{-1.6}$,
    $\theta_{13}=8.50^{+0.20}_{-0.21}$, with the CP violating oscillation phase $\delta=-54^{+39}_{-70}$,
    where the errors represent the one sigma range.
    } \label{angles}
\end{figure}

The present status of neutrino physics is summarised in Figs.\ref{mass},\ref{angles}.
Despite the great pace of progress in neutrino physics, there are still several unanswered experimental questions, as follows:
\begin{itemize}
\item Is the atmospheric neutrino angle $\theta_{23}$ in the first or second octant?
\item Do neutrino mass squared eigenvalues have a normal ordering (NO) or inverted ordering (IO)?
\item What is the value of the lightest neutrino mass?
\item Are neutrinos Dirac or Majorana?
\item Is \CP violated in the leptonic sector and if so by how much?
\end{itemize}
What is the \CP violating phase $\delta$? Is the current hint $\delta \sim -\pi/2$
going to hold up? It is common but incorrect 
to refer to the mass squared ordering question as the
``neutrino mass hierarchy''. However the ``ordering'' question
is separate from whether neutrinos are
hierarchical in nature or approximately degenerate,
which is to do with the lightest neutrino mass.
There are many neutrino experiments underway or planned which will address these questions \cite{expts}.

There are further questions about neutrinos one might ask in the context of the
flavour problem as a whole:
\begin{itemize}
\item What is the origin of the neutrino mass?
\item Why are neutrino masses so tiny compared to charged fermion masses? 
\item Why are at least two neutrino masses not very hierarchical?
\item Why are PMNS mixing angles large?
\item What is the origin of \CP violation?
\end{itemize}

\section{Lepton Mixing}
\label{patterns}

\subsection{Tri-bimaximal Mixing}

A simple pattern of lepton mixing which came to dominate the model building community until the measurement of the reactor angle is the tribimaximal (TB) mixing matrix \cite{Harrison:2002er}. 
It predicts zero reactor angle $\theta_{13}=0$,  maximal atmospheric angle $s^2_{23}=1/2$, 
or $\theta_{12}=45^\circ$, and 
a solar mixing angle given by $s_{12}=1/\sqrt{3}$,
i.e. $\theta_{12}\approx 35.3^\circ$. The mixing matrix is given explicitly by
\begin{equation}\label{TB}
U_{\rm TB} =
\left(
\begin{array}{ccc}
\sqrt{\frac{2}{3}} &  \frac{1}{\sqrt{3}}
&  0 \\ - \frac{1}{\sqrt{6}}  & \frac{1}{\sqrt{3}} &  \frac{1}{\sqrt{2}} \\
\frac{1}{\sqrt{6}} & -\frac{1}{\sqrt{3}} &  \frac{1}{\sqrt{2}}    
\end{array}
\right).
\end{equation}%

\subsection{Deviation parameters}
After the measurement of the reactor angle, TB mixing is excluded. However, TB mixing 
still remains a reasonable approximation to lepton mixing for the solar and atmospheric angles.
It therefore makes sense to expand the angles about their TB values \cite{King:2007pr,Pakvasa:2008zz}:
\begin{eqnarray}
\sin\theta_{12} &=& \frac{1}{\sqrt{3}} (1+s),\\
\sin\theta_{23} &=& \frac{1}{\sqrt{2}} (1+a),\\
\sin\theta_{13} &=& \frac{r}{\sqrt{2}},
\end{eqnarray}
where $s,\,a$, and $r$ are the ($s$)olar, ($a$)tmospheric and ($r$)eactor deviation parameters
such that TB mixing  \cite{Harrison:2002er} is recovered for $s=a=r=0$. For example, TBC mixing 
\cite{King:2012vj} corresponds to $s=a=0$
and $r=\theta_C$, where $\theta_C$ is the Cabibbo angle, which is consistent with data at three sigma.
Certain mixing schemes give correlations between these parameters and $\cos \delta$, as discussed in \cite{King:2007pr,Ballett:2013wya}.
TM1 mixing where the first column
of the TB matrix in Eq.\ref{TB} is preserved gives $a=r\cos \delta$.
TM2 mixing where the second column
of the TB matrix in Eq.\ref{TB} is preserved gives $a=-\frac{1}{2}r\cos \delta$.
If TB mixing is corrected by small charged lepton mixing we have $s=r\cos \delta$,
the ``solar sum rule''
\cite{King:2005bj,King:2007pr}.

\section{The Seesaw Playground}

\subsection{Seesaw mechanism with one right-handed neutrino}
Let us first summarize
the different types of neutrino mass that are possible.
There are Majorana masses of the form
\begin{equation}
\mathcal{L}^{LL}_\nu = 
-\frac{1}{2}m^{\nu}_{LL} \overline{{\nu}_{L}} {\nu}_{L}^{c} + \mathrm{H.c.}
\label{mLL}
\end{equation}
where $\nu_L$ is a left-handed neutrino field and $\nu_L^c$ is
the CP conjugate of a left-handed neutrino field, in other words
a right-handed antineutrino field.

Such Majorana masses are possible below the electroweak symmetry breaking scale 
since both the neutrino and the antineutrino
are electrically neutral and so
Majorana masses are not forbidden by electric charge conservation.
By contrast, a Majorana mass for the electron would
be strictly forbidden. However such Majorana neutrino masses
violate lepton number conservation, and above the electroweak symmetry 
breaking scale where the SM gauge group is unbroken, 
assuming only the simplest Higgs bosons are present, are
forbidden.
The idea of the simplest version of the see-saw mechanism is to assume
that such terms are zero to begin with, but are generated effectively,
after right-handed neutrinos are introduced \cite{seesaw}.

If we introduce right-handed neutrino fields then there are two sorts
of additional neutrino mass terms that are possible. There are
additional Majorana masses of the form
\begin{equation}
\mathcal{L}^{RR}_\nu = -\frac{1}{2} M_{R} \overline{\nu_R^c} \nu_{R} + \mathrm{H.c.} 
\label{MRR}
\end{equation}
where $\nu_R$ is a right-handed neutrino field and $\nu_R^c$ is
the CP conjugate of a right-handed neutrino field, in other words
a left-handed antineutrino field. In addition there are
Dirac masses of the form, using a different notation for the Dirac neutrino masses,
$m^D\equiv m_{LR}$,
\begin{equation}
\mathcal{L}^{LR}_\nu = -m^D\overline{\nu_L}\nu_R + \mathrm{H.c.} .
\label{mLR}
\end{equation}
Such Dirac mass terms conserve lepton number, and are not forbidden
by electric charge conservation even for the charged leptons and
quarks. Dirac mass terms are generated in the SM from the Yukawa couplings to a Higgs doublet,
$H_u$,
\begin{equation}
\mathcal{L}^{yuk} =- H_u Y^{\nu}\overline L \nu_{R} + \mathrm{H.c.}
\label{yuk}
\end{equation}
with the Dirac mass matrix given by $m^D=v_uY^{\nu}$ where $v_u= \langle H_u \rangle$.

\begin{figure}[t]
\centering
\includegraphics[width=0.80\textwidth]{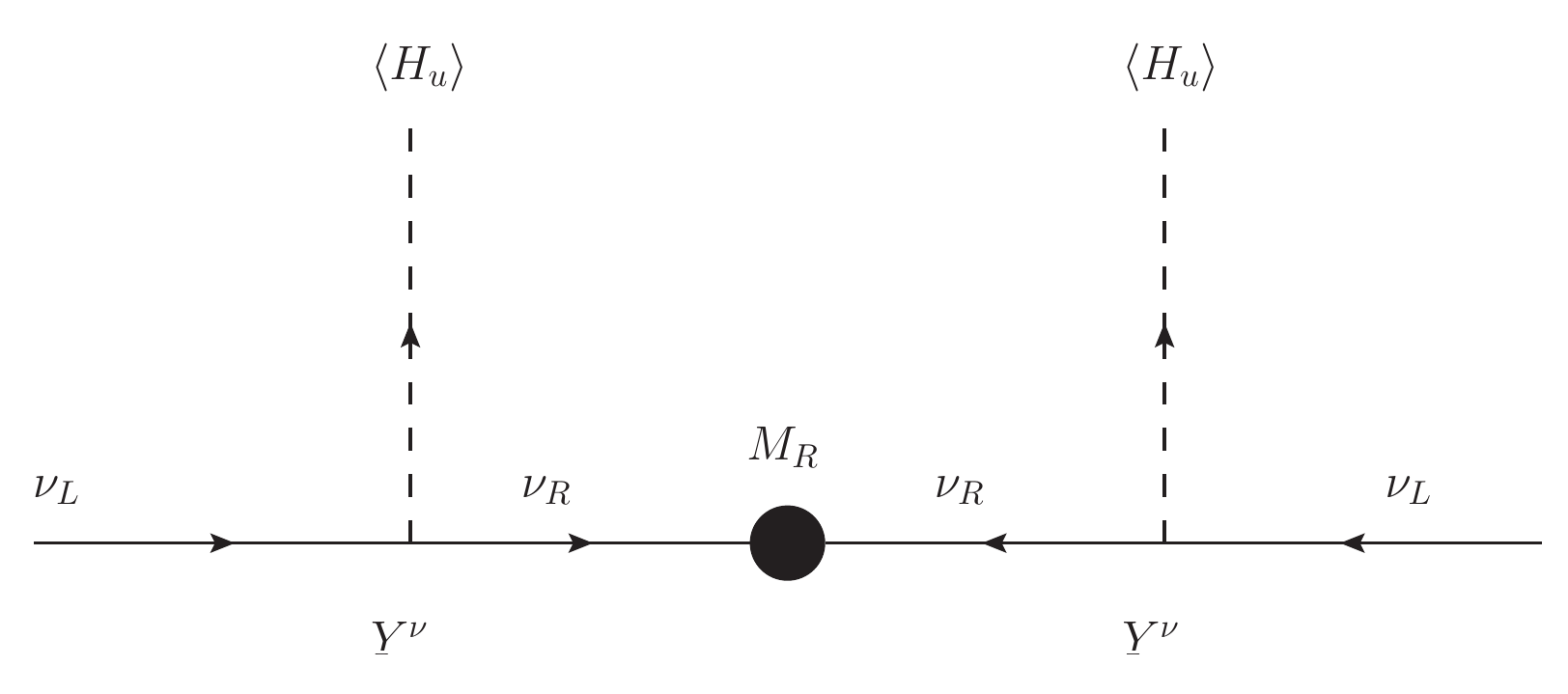}
    \caption{The seesaw mass insertion diagram responsible for the light effective left-handed Majorana neutrino mass 
    $m^{\nu}=-m^DM_{R}^{-1}(m^D)^T$ where the Dirac neutrino mass is $m^D=Y^{\nu}\langle H_u \rangle $.    } \label{seesawfig}
\end{figure}

With the types of neutrino mass discussed
in Eqs.\ref{MRR},\ref{mLR} (but not Eq.\ref{mLL} since we
assume no Higgs triplets) we have the see-saw mass matrix
\begin{equation}
\left(\begin{array}{cc} \overline{\nu_L} & \overline{\nu^c_R}
\end{array} \\ \right)
\left(\begin{array}{cc}
0 & m^D\\
(m^D)^T & M_{R} \\
\end{array}\right)
\left(\begin{array}{c} \nu_L^c \\ \nu_R \end{array} \\ \right)
\label{matrix}
\end{equation}
Since the right-handed neutrinos are electroweak singlets
the Majorana masses of the right-handed neutrinos $M_{RR}$
may be orders of magnitude larger than the electroweak
scale. In the approximation that $M_{RR}\gg m_{LR}$
the matrix in Eq.\ref{matrix} may be diagonalised to
yield effective Majorana masses of the type in Eq.\ref{mLL},
\begin{equation}
m^{\nu}=-m^DM_{R}^{-1}(m^D)^T,
\label{seesaw}
\end{equation}
where we drop the subscript $LL$ on the effective neutrino mass for brevity.
The seesaw mechanism formula is represented by the mass insertion diagram in Fig.\ref{seesawfig}.

The effective left-handed Majorana mass $m^{\nu}$ is naturally
suppressed by the heavy scale $M_{R}$.
If we take $m^D=M_W=80$ GeV
and $M_{R}=M_{GUT}=10^{16}$ GeV
then we find $m^{\nu}\sim 10^{-3}$ eV which looks good for solar
neutrinos.
Atmospheric neutrino masses would require
a right-handed neutrino with a mass below the GUT scale,
as in the single right-handed neutrino model  \cite{King:1998jw}.

\subsection{See-saw mechanism with two right-handed neutrinos }

In this subsection we consider the high scale (classic) {\em see-saw} neutrino model involving just 
{\em two} right-handed neutrinos.
We follow the notation of \cite{King:1999mb}, where the first 
phenomenologically viable model with
two right-handed neutrinos was proposed.
Subsequently
two right-handed neutrino models with two texture zeros were discussed
in \cite{Frampton:2002qc}, however such two texture zero 
models are now phenomenologically excluded \cite{Harigaya:2012bw} for the case of a normal neutrino mass hierarchy considered here. However the original one texture zero case with two right-handed neutrinos
\cite{King:1999mb} remains viable.

The two right-handed neutrinos 
$\nu^{\rm sol}_R$ and $\nu^{\rm atm}_R$
have Yukawa couplings \cite{King:1999mb},
\begin{equation}
\mathcal{L}^{yuk} =(H_u/v_u)(a\overline L_e+ b\overline L_{\mu} +c\overline L_{\tau}){\nu^{\rm sol}_R}
+ (H_u/v_u)(d\overline L_e+ e\overline L_{\mu} +f\overline L_{\tau}){\nu^{\rm atm}_R}+H.c.,
\end{equation}
where $H_u$ is a Higgs doublet
and $v_u$ its vacuum expectation value (VEV).
The heavy right-handed Majorana masses are,
\begin{equation}
\mathcal{L}^{RR}_\nu = M_{\rm sol}\overline{\nu^{\rm sol }_R}({\nu^{\rm sol}_R})^c
+M_{\rm atm}\overline{\nu^{\rm atm }_R}({\nu^{\rm atm}_R})^c +H.c..
\end{equation}
In the basis, with rows
$(\overline \nu_{eL}, \overline \nu_{\mu L}, \overline \nu_{\tau L})$ and columns ${\nu^{\rm atm}_R}, {\nu^{\rm sol}_R}$, the resulting Dirac mass matrix is,
\begin{equation}
m^D=
\left( \begin{array}{cc}
d & a \\
e & b \\
f & c
\end{array}
\right),\ \ \ \ 
(m^D)^T=
\left( \begin{array}{ccc}
d & e & f\\
a & b& c
\end{array}
\right)
\label{mD}
\end{equation}

The (diagonal) 
right-handed neutrino heavy Majorana mass matrix $M_{R}$
with rows $(\overline{\nu^{\rm atm}_R}, \overline{\nu^{\rm sol}_R})^T$ and columns $(\nu^{\rm atm}_R, \nu^{\rm sol}_R)$
is,
\begin{equation}
M_{R}=
\left( \begin{array}{cc}
M_{\rm atm} & 0 \\
0 & M_{\rm sol}
\end{array}
\right),\ \ \ \ 
M^{-1}_{R}=
\left( \begin{array}{cc}
M^{-1}_{\rm atm} & 0 \\
0 & M^{-1}_{\rm sol}
\end{array}
\right)
\label{mR}
\end{equation}

The see-saw formula in Eq.\ref{seesaw} \cite{seesaw} is now interpreted in a matrix sense,
\begin{equation}
m^{\nu}=-m^DM^{-1}_{R}(m^D)^T,
\label{seesaw2}
\end{equation}
where $m^{\nu}$ is the 
the light effective left-handed Majorana neutrino mass matrix
(i.e. the physical neutrino mass matrix),
$m^D$ is the Dirac mass matrix in LR convention and $M_R$ is the (heavy) Majorana
mass matrix. 
Using the see-saw formula 
dropping the overall minus sign which is physically irrelevant,
the light effective left-handed Majorana neutrino mass matrix
$m^{\nu}$
(i.e. the physical neutrino mass matrix) is, by multiplying the matrices in Eqs.\ref{mD},\ref{mR},
\begin{equation}
m^{\nu}=m^DM^{-1}_{R}(m^D)^T=
\left( \begin{array}{ccc}
\frac{a^2}{M_{\rm sol}}+ \frac{d^2}{M_{\rm atm}}& \frac{ab}{M_{\rm sol}}+ \frac{de}{M_{\rm atm}} 
& \frac{ac}{M_{\rm sol}}+ \frac{df}{M_{\rm atm}}  \\
\frac{ab}{M_{\rm sol}}+ \frac{de}{M_{\rm atm}} & \frac{b^2}{M_{\rm sol}}+ \frac{e^2}{M_{\rm atm}}  
& \frac{bc}{M_{\rm sol}}+ \frac{ef}{M_{\rm atm}}  \\
\frac{ac}{M_{\rm sol}}+ \frac{df}{M_{\rm atm}} 
& \frac{bc}{M_{\rm sol}}+ \frac{ef}{M_{\rm atm}}
& \frac{c^2}{M_{\rm sol}}+ \frac{f^2}{M_{\rm atm}}  
\end{array}
\right)
\label{2rhn}
\end{equation}

\subsection{Seesaw with three right-handed neutrinos and sequential dominance}
\begin{figure}[t]
\centering
\includegraphics[width=0.4\textwidth]{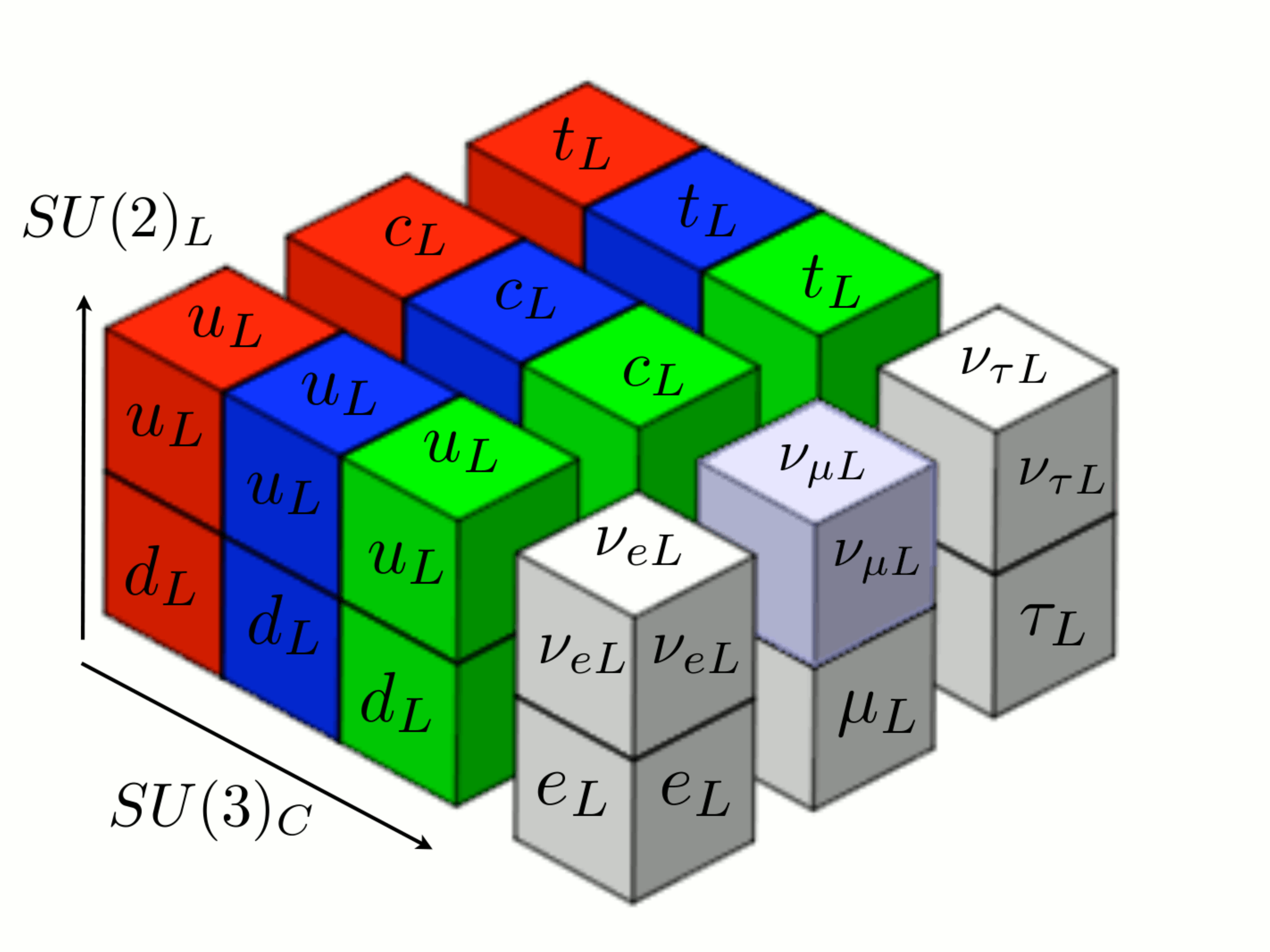}
\includegraphics[width=0.4\textwidth]{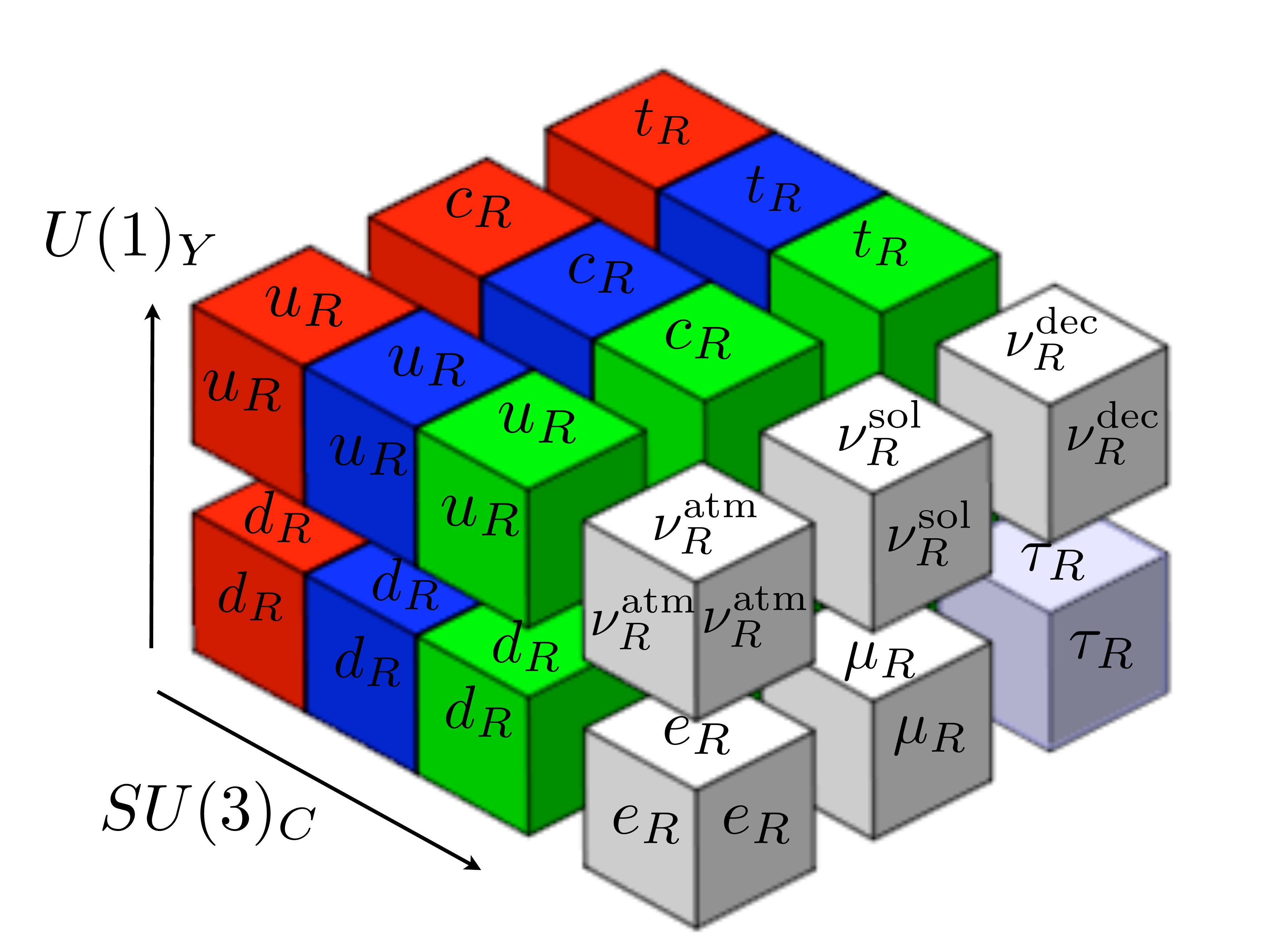}
\vspace*{-4mm}
    \caption{The Standard Model with three right-handed neutrinos defined as
    $(\nu_R^{\rm atm},\nu_R^{\rm sol},\nu_R^{\rm dec})$ which in sequential dominance are mainly responsible for the $m_3,m_2,m_1$ physical neutrino masses, respectively.} \label{SM}
\vspace*{-2mm}
\end{figure}

More generally there may be three right-handed neutrinos,
$\nu^{\rm sol}_R$, $\nu^{\rm atm}_R$ and $\nu^{\rm dec}_R$,
as shown in Fig.\ref{SM}. However, according to sequential dominance \cite{King:1998jw,King:1999mb},
the third right-handed neutrino 
$\nu^{\rm dec}_R$ makes a negligible contribution to the seesaw mechanism,
either due to its high mass or its small Yukawa couplings or both, and so is approximately
decoupled. We are then left with only two right-handed neutrinos $\nu^{\rm sol}_R$ and $\nu^{\rm atm}_R$
as in the two right-handed neutrino model above. 

Motivated by the desire to implement the seesaw mechanism in a natural way,
sequential dominance (SD) \cite{King:1998jw,King:1999mb}
goes further and assumes that the two right-handed neutrinos
$\nu^{\rm sol}_R$ and $\nu^{\rm atm}_R$ have couplings 
$d\ll e,f$ and 
\begin{equation}
 \frac{(e,f)^2}{M_{\rm atm}} \gg \frac{(a,b,c)^2}{M_{\rm sol}}.
\label{SD0}
\end{equation}
By explicit calculation, using Eq.\ref{2rhn}, one can check that 
in the two right-handed neutrino limit $\det m^{\nu} = 0$.
Since the determinant of a Hermitian matrix is the product of mass eigenvalues 
$$
\det (m^{\nu}m^{{\nu}\dagger}) = m_1^2m_2^2m_3^2,
$$
one may deduce that one of the mass eigenvalues of the complex symmetric matrix above
is zero, which under the SD assumption is the lightest one $m_1=0$
with $m_3\gg m_2$ since the model approximates to a single right-handed neutrino model 
\cite{King:1998jw}.
Hence we see that {\it SD implies a normal neutrino mass hierarchy.}
Including the solar right-handed neutrino as a perturbation, it can be shown that,
for $d=0$, together with the assumption of a dominant atmospheric right-handed neutrino 
in Eq.\ref{SD0}, leads to the approximate results for the solar and atmospheric angles 
\cite{King:1998jw,King:1999mb},
\begin{equation}
\tan \theta_{23}\sim \frac{e}{f}, \ \ \ \ \tan \theta_{12} \sim \frac{\sqrt{2}a}{b-c}.
\label{t12}
\end{equation}
Under the above SD assumption, 
each of the right-handed neutrinos contributes uniquely to a particular physical neutrino mass.
The SD framework above with $d=0$
leads to the relations in Eq.\ref{t12} together with the reactor angle bound \cite{King:1999mb},
\begin{equation}
\theta_{13} \lesssim m_2/m_3
\label{13}
\end{equation}
{\it This result shows that SD allows for large values of the reactor angle, consistent with the 
measured value.} Indeed the measured reactor angle, observed a decade after this 
theoretical bound was derived, approximately saturates the upper limit.
In order to understand why this is so, we must go beyond the SD assumptions
stated so far.

\subsection{Playing with the Yukawa couplings on the Seesaw }
\label{CSDn}
Let us 
return to Eq.\ref{mD} and constrain the Yukawa couplings (somehow) to 
$d=0$ and $e=f$, with $b=a$ and $c=-a$ \cite{King:2005bj}.
The motivation is that from Eq.\ref{t12} one then approximately expects 
the good phenomenological relations $t_{23}\sim 1$ and $t_{12}\sim 1/\sqrt{2}$, although the value of the reactor angle bounded by Eq.\ref{13} remains to be seen.
With the above assumption, Eq.\ref{2rhn} becomes
\begin{equation}
m^{\nu}=
\left( \begin{array}{ccc}
\frac{a^2}{M_{\rm sol}} & \frac{a^2}{M_{\rm sol}}
& \frac{-a^2}{M_{\rm sol}}  \\
\frac{a^2}{M_{\rm sol}} & \frac{a^2}{M_{\rm sol}}+ \frac{e^2}{M_{\rm atm}}  
& \frac{-a^2}{M_{\rm sol}}+ \frac{e^2}{M_{\rm atm}}  \\
\frac{-a^2}{M_{\rm sol}} 
& \frac{-a^2}{M_{\rm sol}}+ \frac{e^2}{M_{\rm atm}}
& \frac{a^2}{M_{\rm sol}}+ \frac{e^2}{M_{\rm atm}}  
\end{array}
\right).
\label{CSD}
\end{equation}
By explicit calculation one then finds that the neutrino mass matrix is exactly diagonalised by the TB mixing matrix in Eq.\ref{TB},
\begin{equation}
U_{\mathrm{TB}}^T m^{\nu} U_{\mathrm{TB}}=
\left( \begin{array}{ccc}
0 & 0 & 0  \\
0 & \frac{3a^2}{M_{\rm sol}} & 0 \\
0 & 0 & \frac{2e^2}{M_{\rm atm}}  
\end{array}
\right).
\end{equation}
If the charged lepton mass matrix is diagonal, the interpretation 
is that these constrained couplings $d=0$, $e=f$ with $b=a$ and $c=-a$
lead to TB mixing, with the lightest neutrino mass $m_1=0$, the second lightest neutrino identified
as the solar neutrino with mass $m_2=\frac{3a^2}{M_{\rm sol}}$ and the heaviest neutrino identified
as the atmospheric neutrino with mass $m_3=\frac{2a^2}{M_{\rm atm}}$.
While TB mixing accurately gives the good relations $t_{23}=1$ and $t_{12}=1/\sqrt{2}$,
unfortunately it also gives $\theta_{13}=0$.
This is known as constrained sequential dominance (CSD) \cite{King:2005bj}.

Unfortunately CSD completely fails to saturate the bound
in Eq.\ref{13}, indeed quite the opposite.
However further playing with the Yukawa couplings can lead to the bound
in Eq.\ref{13} being saturated, without messing up the good predictions of CSD for the
solar and atmospheric angles. Indeed one 
can generalise the original idea of CSD to other examples of 
Dirac mass matrix with (in the notation of Eq.\ref{mD})
$d=0$ and $e=f$ as before, but now with 
$b=na$ and  $c=(n-2)a$, for any postive integer $n$. The generalisation is called CSD($n$)
\cite{Antusch:2011ic,King:2013iva,King:2013xba,King:2013hoa,King:2014iia,Bjorkeroth:2014vha}.
The constrained couplings will be justified with the help of discrete family symmetry.
The original CSD in Eq.\ref{CSD} with $b=a$ and $c=-a$ is identified as the 
special case CSD($n=1$).
The motivation for CSD($n$) is that for any $n$ Eq.\ref{t12} implies $t_{23}\sim 1$ and $t_{12}\sim 1/\sqrt{2}$, 
although these results are strongly dependent on
the relative phase between the first and second column
of the Dirac mass matrix. Unfortunately CSD(2) also fails for all choices of phase \cite{Antusch:2011ic},
so the simplest viable case is CSD(3) \cite{King:2013iva}, with CSD(4) also viable
\cite{King:2013xba,King:2013hoa,King:2014iia}.

\subsection{The Littlest Seesaw}

The minimal viable predictive seesaw model corresponds to a two right-handed neutrino model with CSD($3$). In the diagonal charged lepton and right-handed neutrino mass basis
this corresponds to the following Dirac mass matrix
in Eq.\ref{mD} \cite{King:2013iva}:
\begin{equation}
	m^D = 
	Y^{\nu}v_u=\pmatr{0 & a \\ e & 3a \\ e & a }.	\label{mDn}
\end{equation}
These {\it ad hoc} looking couplings may in fact emerge from 
a rather complete SUSY GUT of Flavour based on $A_4\times Z_9\times SU(5)$~\cite{Bjorkeroth:2015ora}.
Here we simply assume these couplings motivated by the desire to obtain an approximately 
maximal atmospheric angle $\tan \theta_{23} \sim e/f \sim 1$ and 
trimaximal solar angle 
$\tan \theta_{12} \sim \sqrt{2}a/(b-c) \sim 1/\sqrt{2}$.
Since experiment indicates that the bound $\theta_{13} \lesssim m_2/m_3$ is almost saturated, these schemes require certain phase choices 
$\arg (a/e)$ in order to achieve the desired reactor angle, leading to predictions for the \CP-violating phase $\delta_{CP}$, discussed below.%

The low energy effective Majorana neutrino mass matrix in Eq.\ref{2rhn}
in the two right-handed neutrino case may be written as,
\begin{equation}
	m^\nu = m_a 
	\left(
\begin{array}{ccc}
	0&0&0\\0&1&1\\0&1&1 
	\end{array}
\right)
	+ m_b e^{i\eta} 
	\left(
\begin{array}{ccc}
	1&3&1\\3&9&3\\1&3&1
	\end{array}
\right),
	\label{eq:mnu2}
\end{equation}
where $\eta$ is the only physically important phase, which depends on the relative phase between the first and second column of the Dirac mass matrix, $\arg (a/e)$.
By comparing Eqs.\ref{2rhn} and \ref{eq:mnu2} 
we identify $m_a=\frac{e^2}{M_{\rm atm}}$ and $m_b=\frac{a^2}{M_{\rm sol}}$.
This can be thought of as the minimal (two right-handed neutrino) predictive seesaw model since
only three parameters $m_a, m_b, \eta$ describe the entire neutrino sector (three neutrino masses
and the PMNS matrix).
CSD($3$) with two right-handed neutrinos always 
predicts the lightest physical neutrino mass to be zero, $m_1=0$.
One can also check that 
\begin{equation}
m^\nu
\left(
\begin{array}{c}
2 \\
-1\\
1
\end{array}
\right)
=
\left(
\begin{array}{c}
0 \\
0\\
0
\end{array}
\right).
\label{CSD(n)a}
\end{equation}
In other words the column vector $(2,-1,1)^T$
is an eigenvector of $m^\nu $ with a zero eigenvalue, i.e. it is the first column of the PMNS mixing matrix,
corresponding to $m_1=0$,
which means so called TM1 mixing in which the first column of the TB mixing matrix in Eq.\ref{TB}
is preserved, while the other two columns are allowed to differ (in particular the reactor angle will be non-zero).

The numerical predictions are given in Table \ref{tab:model} for some optimal choice 
of input parameters, where they are compared to the global best fit values 
from \cite{Gonzalez-Garcia:2014bfa} (setting $m_1=0$). The model in \cite{Bjorkeroth:2015ora}
predicts that the phase $\eta$ is one of the ninth roots of unity, and we have selected 
$\eta = 2\pi/3$.
The agreement between CSD(3) and data is 
within about one sigma for all the parameters.

\begin{table}[ht]
\renewcommand{\arraystretch}{1.2}
\centering
\begin{tabular}{| c c c | c c c c c c c |}
\hline
\rule{0pt}{4ex}%
\makecell{$m_a$ \\ {\scriptsize (meV)}} & \makecell{$m_b$ \\ {\scriptsize (meV)}} & 
\makecell{$\eta$  \\ {\scriptsize (rad)}}  	& \makecell{$\theta_{12}$ \\ {\scriptsize ($^{\circ}$)}} & \makecell{$\theta_{13}$ \\ {\scriptsize ($^{\circ}$)}}  & \makecell{$\theta_{23}$ \\ {\scriptsize ($^{\circ}$)}} & \makecell{$\delta_{\mathrm{CP}}$ \\ {\scriptsize ($^{\circ}$)}} & \makecell{$m_1$ \\ {\scriptsize (meV)}}
& \makecell{$m_2$ \\ {\scriptsize (meV)}} & \makecell{$m_3$ \\ {\scriptsize (meV)}} \\ [2ex] \hline 
\rule{0pt}{4ex}%
26.57		& 2.684		& $ \dfrac{2\pi}{3} $	& 34.3		& 8.67		& 45.8		& -86.7		& 0 & 8.59		& 49.8 \\[1.7ex]
\hline
Value & from & \cite{Gonzalez-Garcia:2014bfa} & 33.48$^{+0.78}_{-0.75}$ & 8.50$^{+0.20}_{-0.21}$ &
42.3$^{+3.0}_{-1.6}$  &  -54$^{+39}_{-70}$ & 0 & 8.66$\pm 0.10$ & 49.57$\pm 0.47$ \\
\hline
\end{tabular}
\caption{Parameters and predictions for CSD(3) with a fixed phase $ \eta = 2\pi/3 $
from \cite{Bjorkeroth:2015ora}.
These predictions may be compared to the global best fit values 
from \cite{Gonzalez-Garcia:2014bfa} (for $m_1=0$), given on the last line.}
\label{tab:model}
\end{table}

\subsection{The Littlest Leptogenesis}

Using the results in Table \ref{tab:model}, the baryon asymmetry of the Universe
(BAU) resulting from $N_1 = N_{\rm atm}$ leptogenesis was estimated for this model \cite{Bjorkeroth:2015tsa}:
\begin{equation}
	Y_B \approx 2.5 \times 10^{-11}\sin \eta \left[\frac{M_1}{10^{10} ~\mathrm{GeV}} \right].
\label{BAU}
\end{equation}
Using $\eta = 2\pi/3$ and the observed value of $ Y_B $ fixes the lightest right-handed neutrino mass:
\begin{equation}
	M_1 = M_{\rm atm}  \approx 3.9 \times 10^{10} ~\mathrm{GeV}.
\end{equation}
Note that the phase $\eta$ controls the BAU via leptogenesis in Eq.\ref{BAU}.
The phase $\eta$ also controls the entire PMNS matrix, including all the lepton mixing angles as well as all low energy \CP violation.
The single phase $\eta$ is the therefore the source of all \CP violation in this model,
including both \CP violation in neutrino oscillations and in leptogenesis, providing a direct link between these two phenomena in this model. We not only have a correlation between the sign of the BAU and the sign of 
low energy leptonic \CP violation, but we actually know the value of the 
leptogenesis phase: it is $\eta = 2\pi/3$ which leads to the observed excess of matter over antimatter
for $M_1 \approx 4.10^{10}$ GeV
together with an observable neutrino oscillation phase $\delta_{\mathrm{CP}} \approx -\pi/2$.

\section{Conclusion}
\label{conclusion}
Although the discovery of neutrino mass and mixing does not ``prove'' the
validity of the seesaw mechanism any more than the discovery of proton decay would ``prove'' GUTs,
it suggests it as a simple and attractive possibility.
However with many undetermined input parameters, {\it a priori}
it is far from clear how to implement the seesaw in order to make it into a predictive framework.

We have focused on natural implementations of the seesaw mechanism in which large cancellations do not occur. In one such natural framework, 
one of the right-handed neutrinos is dominantly responsible
for the atmospheric neutrino mass, while a second right-handed neutrino accounts for the solar neutrino mass and a third right-handed neutrino is approximately 
decoupled, leading to an effective two right-handed neutrino model. 
The main predictions of such sequential dominance
(SD) are Majorana neutrinos with a normal neutrino mass hiearchy and a reactor angle satisfying the bound $\theta_{13} \lesssim m_2/m_3$, suggesting a large reactor angle
a decade before it was measured. 

In order to understand why the reactor angle bound is approximately saturated,
we need to play with the Yukawa couplings.
The original idea of constrained sequential dominance (CSD) provides a good explanation
of the tri-bimaximal solar and atmospheric angles but unfortunately
predicts a zero reactor angle. However, following a further unsuccessful attempt,
a viable scheme was found called, CSD(3), affectionately dubbed here the ``Littlest Seesaw''.

The Littlest Seesaw, with two right-handed neutrinos,
gives a successful
desciption of neutrino mass and the PMNS matrix in terms of just three input parameters
in Eq.\ref{eq:mnu2}.
It predicts a normally ordered and very hierarchical neutrino
mass spectrum with the lightest neutrino mass being zero and both
atmospheric angle and leptonic 
\CP violation being close to maximal, as shown in Table~\ref{tab:model}.

The Littlest Seesaw can be derived from 
a combination of GUT and family symmetry, involving a discrete family symmetry $A_4\times Z_9$
together with an $SU(5)$ SUSY GUT. 
The $A_4$ provides the vacuum aligments
responsible for the otherwise {\it ad hoc} Yukawa couplings, while the $Z_9$ fixes
the \CP violating input phase $\eta$ to be one of the ninth roots of unity, which is selected to be
$\eta = 2\pi/3$. Indeed $\eta$ is the only source of \CP violation in the model,
and is responsible for both the oscillation phase and leptogenesis, providing the most direct link
possible between these two phenomena.

In conclusion, although the origin of neutrino mass is unknown, one of the minimal possibilities is the 
see-saw mechanism with heavy right-handed neutrinos. 
In the absence of any other new physics, the seesaw mechanism will surely continue to provide a playground for theorists.

\vspace{0.1in}
SFK acknowledges partial support from the STFC Consolidated ST/J000396/1 grant and 
the European Union FP7 ITN-INVISIBLES (Marie Curie Actions, PITN-
GA-2011-289442).

\end{document}